\documentclass[amsmath,amssymb,twocolumn,showpacs]{revtex4}

\usepackage{mathptmx}

\usepackage{graphicx}
\usepackage{bm}
\usepackage{amsthm}

\newcommand{\R}{\mathbb{R}}
\newcommand{\Z}{\mathbb{Z}}

\theoremstyle{plain}
\newtheorem{theorem}{Theorem}

\begin{document}

\title{Counterintuitive ground states in soft-core models}

\author{Henry Cohn}
\affiliation{Microsoft Research New England, One Memorial Drive,
Cambridge, Massachusetts 02142, USA} \email{cohn@microsoft.com}

\author{Abhinav Kumar}
\affiliation{Department of Mathematics, Massachusetts Institute of
Technology, Cambridge, Massachusetts 02139, USA}
\email{abhinav@math.mit.edu}

\date{November 7, 2008}

\begin{abstract}
It is well known that statistical mechanics systems exhibit subtle
behavior in high dimensions.  In this paper, we show that certain
natural soft-core models, such as the Gaussian core model, have
unexpectedly complex ground states even in relatively low dimensions.
Specifically, we disprove a conjecture of Torquato and Stillinger, who
predicted that dilute ground states of the Gaussian core model in
dimensions $2$ through $8$ would be Bravais lattices.  We show that in
dimensions $5$ and $7$, there are in fact lower-energy non-Bravais
lattices.  (The nearest three-dimensional analog is the hexagonal
close-packing, but it has higher energy than the face-centered cubic
lattice.) We believe these phenomena are in fact quite widespread, and
we relate them to decorrelation in high dimensions.
\end{abstract}


\maketitle

\section{Introduction}

One of the most natural soft-core models in statistical mechanics is
the Gaussian core model (introduced by Stillinger \cite{S}), in which
identical particles interact via a repulsive Gaussian pair potential.
This is not only a beautiful theoretical model, but also a reasonable
model for the effective interaction (via entropic repulsion) between
the centers of mass of two polymers, namely, the Flory-Krigbaum
potential \cite{FK,LBHM}.  Much work has gone into characterizing the
phase diagram and ground states of the Gaussian core model
\cite{LLWL,PSG}.

We use the Gaussian core model as a test case for studying the
emergence of long-range structure in classical ground states.  In two
or three dimensions, these ground states are typically lattices, and
even Bravais lattices.  The theory behind this phenomenon is poorly
understood: the Lennard-Jones potential in two dimensions has been
rigorously analyzed by Theil \cite{T}, and S\"ut\H o \cite{Su1,Su2} has
analyzed potential functions whose Fourier transforms are nonnegative
and have compact support, but for no purely repulsive soft-core
potential in more than one dimension is there a compelling argument for
crystallization (let alone a proof). In the present paper, we show the
subtlety of this problem by exhibiting counterintuitive ground states
with different structure than anticipated.

Specifically, we study the Gaussian core model for dilute systems in
high-dimensional spaces.  Although that may sound arcane, such systems
play an important role in statistical physics.  First, they include
sphere packing problems as a limiting case.  Packing in high dimensions
is of fundamental importance in communication and information theory,
because (as Shannon discovered) finding codes for efficient
communication in the presence of noise amounts to a packing problem in
the high-dimensional space of possible signals.

Second, such systems provide an intriguing test case for the
decorrelation effect, a fundamental phenomenon predicted by Torquato
and Stillinger \cite{TS1}: in loose terms, unconstrained spatial
correlations should vanish asymptotically in high dimensions, and all
multibody correlations will be reducible to the pair correlation
function. Although it seems difficult to justify rigorously,
decorrelation leads to surprising conjectures such as the existence of
extraordinarily dense disordered packings in high dimensions (with
important implications in information theory).  See also Ref.\
\cite{PZ} for a replica symmetry-breaking approach to amorphous
packings in high dimensions.

This line of reasoning suggests that glassy states of matter are
intrinsically more stable than crystals in high dimensions, which
stands in stark contrast to intuition derived from most two- or
three-dimensional systems.  In three dimensions, for example, the
low-density ground state for the Gaussian core model is the
face-centered cubic (fcc) lattice, which has lower energy than the
competing hexagonally close-packed (hcp) lattice, let alone disordered
structures. In the present paper, we show that the opposite happens in
as few as five dimensions: relatively exotic non-Bravais lattices
improve on more familiar structures.  To find this behavior in such a
low dimension is unexpected, and while we cannot demonstrate the full
decorrelation effect (for example, with completely amorphous packings),
our results show that the role of order and structure in even
low-dimensional ground states is more subtle than was previously
realized.

Our direct motivation is a recent prediction by Torquato and Stillinger
\cite{TS2} for the ground states of the Gaussian core model in
moderately high dimensions (up through $\R^8$ and also $\R^{24}$).
Specifically, at sufficiently low particle density, they conjectured
that the ground states are the Bravais lattices corresponding to the
densest known sphere packings, and at sufficiently high particle
density they conjectured that the ground states were the reciprocal
Bravais lattices.  In the case of $\R^2$, $\R^8$, and $\R^{24}$, this
agrees with an earlier conjecture of Cohn and Kumar (Conjecture 9.4 in
Ref.\ \cite{CK}).  Zachary, Stillinger, and Torquato \cite{ZST} have
given strong numerical evidence that these are indeed the true ground
states among known families of Bravais lattices. However, in this paper
we disprove Torquato and Stillinger's conjecture by exhibiting
non-Bravais lattices with lower energy in the low density limit in
$\R^5$ and $\R^7$.

These improved lattices in fact correspond to tight sphere packings
(i.e., sphere packings that are not only as dense as possible globally,
but also locally, in the sense that there are no missing spheres, small
gaps, etc.). Conway and Sloane \cite{CS} provided a conjecturally
complete list of tight packings in low dimensions, and our ground
states can be found in their list.  They stand in the same relationship
to the optimal Bravais lattices as the hcp packing stands to the fcc
packing in $\R^3$, but the energy comparisons work out notably
differently. This is in effect another facet of decorrelation. Even
within the restrictive class of tight packings, in high dimensions
Bravais lattices are no longer energetically favored.  Instead,
somewhat less regular structures are preferred.

For comparison to the mathematical literature (and, in particular,
Ref.\ \cite{CS}), note that mathematicians use ``lattice'' to mean
``Bravais lattice'' and ``periodic packing'' to mean ``lattice with a
basis.'' In this paper, we follow the physics terminology.

\section{Theta series}

All of our work in this paper takes place in the low density limit.
Because of the scaling invariance of Euclidean space, we can instead
fix the particle density and rescale the Gaussian. Specifically, we use
the potential function $V(r) = e^{-\alpha r^2}$ between two particles
at distance $r$, and we let $\alpha$ tend to infinity, which
corresponds to taking the low-density limit.

The \emph{theta series} for a packing $\mathcal{P}$ (i.e., a collection
of particle locations) is a generating function that describes the
average number of particles at a given distance from a particle in
$\mathcal{P}$. Specifically,
$$
\Theta_{\mathcal{P}}(q) = \sum_r N_r \,q^{r^2},
$$
where the sum is over all distances $r$ between points in the packing,
$N_r$ denotes the average over all $x \in \mathcal{P}$ of the number of
$y \in \mathcal{P}$ such that $|x-y|=r$, and $q$ is a formal variable.
The use of $r^2$ rather than $r$ in the exponent is traditional in
mathematics.  Note that the theta series encodes the same information
as the pair correlation function; we use this notation since it is
convenient for the Gaussian core model.

Under the Gaussian core model potential function $V(r) = e^{-\alpha
r^2}$, the average energy per point in $\mathcal{P}$ equals
$(\Theta_{\mathcal{P}}\big(e^{-\alpha}\big)-1)/2$.  (We subtract $1$ to
correct for the $r=0$ term in the theta series, which would correspond
to a self-interaction, and we divide by $2$ to avoid double counting.)
Thus, computing theta series is exactly the same as computing energy in
the Gaussian core model. The limit as $\alpha \to \infty$ of energy
corresponds to the limit as $q \to 0$ of the theta series.

Given two packings with the same density (i.e., the same number of
particles per unit volume in space), we can easily compare their
behavior in the $q \to 0$ limit.  Suppose their theta series are
$$
\Theta_1 = 1 + a_{r_1} q^{r_1^2} + a_{r_2} q^{r_2^2} + \cdots
$$
and
$$
\Theta_2 = 1 + b_{s_1} q^{s_1^2} + b_{s_2} q^{s_2^2} + \cdots
$$
with $r_1 < r_2 < \cdots$ and $s_1 < s_2 < \cdots$.  To compare
$\Theta_1$ with $\Theta_2$, we need only consider the first term at
which they differ.  If $r_1 > s_1$, then $\Theta_1 < \Theta_2$ for
small $q$; if $r_1=s_1$, then the comparison amounts to whether
$a_{r_1} < b_{s_1}$.  If $r_1=s_1$ and $a_{r_1} = b_{s_1}$, then we
must proceed to the next term.

Corresponding to any point configuration in $\R^n$, we obtain a sphere
packing by centering identical spheres at the points of the
configuration, with the maximal possible radius subject to avoiding
overlap.  The density of the packing is the fraction of space covered.
To avoid confusion, we will distinguish between the \emph{particle
density} (the number of particles per unit volume in space) and the
\emph{packing density} (the fraction of space covered by balls).

As pointed out above, maximizing packing density is a consequence of
minimizing energy in the Gaussian core model in the $\alpha \to \infty$
limit (with fixed particle density): the dominant contribution to the
Gaussian energy comes from the smallest distance between points, which
is large exactly when the packing density is large.  In other words,
the problem of maximizing the sphere packing density arises naturally
as the low-density limit of the Gaussian core model.

\section{Tight packings}

In most dimensions, the sphere packing problem exhibits high
degeneracy, in the sense that there are many geometrically distinct,
optimal solutions (such as in three dimensions, with the fcc and hcp
packings and their relatives). Conway and Sloane \cite{CS} gave a
conjectural classification of all the tight packings in low dimensions.
Here, tight means roughly that the global density is maximized and
furthermore no local changes can add more spheres. (For example,
removing one sphere from a dense packing leaves the global density
unchanged, but the result is no longer tight.)  The precise definition
of tightness in Ref.\ \cite{CS} is problematic; see Ref.\ \cite{K} for
details on the problem and better definitions. Because they recognized
that their definition was only tentative, Conway and Sloane
characterized tightness by articulating ``postulates'' that they felt a
correct definition should satisfy. These postulates are by no means
obvious statements; instead, they are empirical observations from
Conway and Sloane's study of the packing problem.

Conway and Sloane \cite{CS} postulate that, in dimensions up to $8$,
every tight packing fibers over a tight packing whose dimension is the
previous power of $2$.  To say that a packing $\mathcal{P}$ fibers over
$\mathcal{Q}$ means that $\mathcal{P}$ can be decomposed into parallel
layers lying in $\dim(\mathcal{Q})$-dimensional subspaces, each of
which is a packing isometric to $\mathcal{Q}$.  (In fact, in tight
packings of dimensions up to $8$ it will be a translate of
$\mathcal{Q}$.) The locations of these parallel subspaces should
themselves be determined by another tight packing. Although the Conway
and Sloane postulates are only conjectures, they seem likely to be true
and in this paper we assume their truth (but we note which ones are
required for each theorem).

\section{Dimensions up to $4$}

In $\R^1$, there is exactly one tight packing, namely that given by the
integers.  It is provably optimal for the Gaussian core model by
Proposition~9.6 in Ref.\ \cite{CK}.

In $\R^2$, the triangular lattice $A_2$ is the only tight packing.
Montgomery \cite{M} showed that it is optimal among all Bravais
lattices for the Gaussian core model, and it was conjectured in Ref.\
\cite{CK} that it is optimal among all lattices.

In $\R^3$, all tight packings fiber over the triangular lattice $A_2$.
In other words, they are formed by stacking triangular layers, with the
layers nestled together as densely as possible; each additional layer
involves a binary choice for how to place it relative to the previous
layer. These are the Barlow packings (i.e., the stacking variants of
the fcc and hcp packings).  It is not hard to check that, among these
packings, the face-centered cubic lattice minimizes energy in the
Gaussian core model in the low particle-density limit.  This is
consistent with the conjecture in Ref.\ \cite{TS2}.

In $\R^4$, there is only one tight packing, namely the $D_4$ or
checkerboard lattice (it is shown in Ref.\ \cite{CS} that only one
tight packing fibers over $A_2$). It is defined to be the set of all
integral points whose coordinates have even sum:
$$
D_4 = \left\{x \in
\Z^4 : \sum_{i=1}^4 x_i \equiv 0 \!\!\!\!\!\pmod{2}\right\}.
$$
The uniqueness of $D_4$ is remarkable, compared with the diversity of
tight packings in $\R^3$, and the $D_4$ lattice plays a fundamental
role as a building block for higher-dimensional structures. It also
appears that, much like the triangular lattice, $D_4$ may be
universally optimal, in the sense that it is the ground state of the
Gaussian core model at any density.

\section{Dimension $5$}

In $\R^5$, every tight packing fibers over $D_4$, with the distance
between successive layers being $1$. To specify such a packing, one
need only specify how each four-dimensional layer is translated
relative to its neighbors. The \emph{deep holes} in $D_4$ (the points
in space furthest from the lattice) are located at $(1,0,0,0)$,
$(1/2,1/2,1/2,1/2)$, and $(1/2,1/2,1/2,-1/2)$, as well as of course the
translates of these points by vectors in $D_4$. Each layer of a tight
packing in $\R^5$ must either be an untranslated copy of $D_4$ or be
translated by one of these vectors, so that the distance between layers
is minimized; furthermore, adjacent layers must be translated by
different vectors. In other words, the spheres in each layer must be
nestled into the gaps in the adjacent layers.

If we let $a$ denote the translation vector $(0,0,0,0)$, $b$ denote
$(1,0,0,0)$, etc., then each layer must be translated by one of $a$,
$b$, $c$, or $d$, and no two adjacent layers can be translated by the
same vector. In other words, a tight packing in $\R^5$ is specified by
a four-coloring of the integers (if we treat $a$, $b$, $c$, and $d$ as
``colors'').

For example, the $D_5$ packing, which is the Bravais lattice with the
highest packing density, corresponds to the following coloring:
\setlength{\unitlength}{1cm}
\begin{center}
\begin{picture}(5.5,0.8)
\put(0,0.4){$\ldots$}
\put(5.5,0.4){$\ldots$}
\put(1,0.4){\circle{0.5}}
\put(2,0.4){\circle{0.5}}
\put(3,0.4){\circle{0.5}}
\put(4,0.4){\circle{0.5}}
\put(5,0.4){\circle{0.5}}
\put(0.5,-0.1){\makebox(1,1){$a$}}
\put(1.5,-0.1){\makebox(1,1){$b$}}
\put(2.5,-0.1){\makebox(1,1){$a$}}
\put(3.5,-0.1){\makebox(1,1){$b$}}
\put(4.5,-0.1){\makebox(1,1){$a$}}
\end{picture}
\end{center}
Note that the symmetries of the $D_4$ lattice arbitrarily permute $a$,
$b$, $c$, and $d$, so the choice of labeling is irrelevant.  For $D_5$,
all that matters is that the layers alternate between two colors.

Conway and Sloane found that four tight packings are uniform, in the
sense that all spheres play the same role (rather than the less
symmetric situation of having several inequivalent classes of spheres).
In addition to $D_5 = \Lambda_5^1$, the three others correspond to the
following patterns:
\begin{align*}
\Lambda_5^2 &\ :\ \cdots abcdabcd\cdots,\\
\Lambda_5^3 &\ :\ \cdots abcabc \cdots,\\
\Lambda_5^4 &\ :\ \cdots bacbdcadbacbdcad \cdots.
\end{align*}
These three additional lattices are not Bravais lattices, but rather
lattices with bases.

One can calculate that the theta series for $\Lambda_5^1$ is $ 1 +40q^2
+ 90q^4 + 240q^6 + \cdots, $ while the theta series of $\Lambda_5^2$ is
$ 1 + 40 q^2 + 88q^4 + 16q^5 + \cdots. $ It follows that $\Lambda_5^2$
has lower energy  than $D_5$ in the $q \to 0$ limit, which disproves
Torquato and Stillinger's conjecture.  In fact, the situation is even
worse for $D_5$, which is not only suboptimal but in fact the worst
tight packing of all.

\begin{theorem}
Under Postulates 2, 4, and 5 of Ref.\ \cite{CS}, the Bravais lattice
$D_5$ has the highest energy among all the tight five-dimensional
lattices, in the $q \to 0$ limit.
\end{theorem}

To complete this calculation, we require four geometrical facts about
$D_4$.  Specifically, each lattice point has $24$ neighboring lattice
points at squared distance $2$, the next closest lattice points are
$24$ more at squared distance $4$, each deep hole has $8$ neighboring
lattice points at squared distance $1$, and the next closest lattice
points to a deep hole are $32$ points at squared distance $3$.  These
assertions are easily checked by a short calculation.

\begin{proof}
Let $\Lambda$ be a tight five-dimensional lattice, obtained by a
four-coloring of the integers.  We first observe that every sphere in
$\Lambda$ must have $40$ neighbors at squared distance $2$, for the
following reason.  Without loss of generality, we may assume that layer
$0$ is colored $a$ and layer $1$ is colored $b$ (since the different
deep holes are equivalent under the symmetries of $D_4$). Now, layer
$-1$ cannot be colored $a$ either, so the layers $0$, $1$, and $-1$
contribute $24 + 8 + 8 = 40$ neighbors of a given sphere in layer $0$.
(Every sphere in $D_4$ has $24$ neighbors, which accounts for the $24$
from layer $0$, and each deep hole in $D_4$ is at distance $1$ from $8$
points of $D_4$.) Therefore the theta series of $\Lambda$ must start
with $1+40q^2 + \cdots$.

The next smallest possible squared distance in $\Lambda$ is $4$
(squared distance $3$ does not occur in $D_4$, and it cannot occur
between adjacent layers since that would amount to having a lattice
point at squared distance $2$ from a deep hole). There are $24$ spheres
at that distance in $D_4$, and $32$ in each of layers $\pm 1$, for a
total of $88$.  The only way there can be more is if they come from
layers $\pm 2$, and each of those layers contributes one sphere (lying
over the origin) if and only if it is colored the same as layer $0$.
Since $D_5$ corresponds to the coloring $\cdots abababa \cdots$, its
theta function has the maximum contribution to the $q^4$ term, making
it the worst for energy as $q \to 0$. Furthermore, among all tight
lattices only $D_5$ maximizes that term, so it is the unique pessimum.
\end{proof}

The lattice $\Lambda_5^2$ turns out to be the best.

\begin{theorem}
Under Postulates 2, 4, and 5 of Ref.\ \cite{CS}, the lattice
$\Lambda_5^2$ has the lowest energy among all the tight
five-dimensional lattices, in the $q \to 0$ limit.
\end{theorem}

\begin{proof}
The proof is similar to that of the previous theorem. Let $\Lambda$ be
a tight packing as above, fibered over $D_4$. We may assume as before
that layer $0$ is colored $a$. The first two terms of the theta series
of $\Lambda$ are $1$ and $40q^2$. Now, if layer $2$ or layer $-2$ were
colored $a$, then $\Lambda$ would have a larger $q^4$ term than
$\Lambda_5^2$, making it worse for potential energy in the $q \to 0$
limit. Therefore we may assume neither $2$ nor $-2$ is colored $a$. The
theta series is now determined up to the $q^8$ term, and it equals
$1+40q^2 + 88q^4 + 16q^5 + 192q^6 + 64q^7 + 152q^8 + \cdots$.

The $q^9$ term is not yet determined, since it depends on layers $3$
and $-3$.  Merely being three layers apart contributes $3^2$ to the
squared distance, so they contribute to the $q^9$ term if and only if
they are colored $a$.  Thus, to minimize energy they must not be
colored $a$. In other words, two layers of the same color must be
separated by at least $4$. The only way to do this is to color the
layers $\cdots abcdabcd \cdots$, up to permutations of the four colors.
Since permuting the four colors will not change the resulting lattice
(because of the symmetries of $D_4$), we see that $\Lambda_5^2$ is the
unique best lattice among all the tight five-dimensional lattices in
the $q \to 0$ limit.
\end{proof}

\section{Dimension $6$}

In $\R^6$, the way to form tight packings is again to fiber over $D_4$,
and we must use the triangular lattice $A_2$ to arrange the fibers
(with $A_2$ normalized so the closest lattice points are at distance
$1$). Thus, we are looking for four-colorings of the triangular lattice
$A_2$, where the colors specify which translation vector to use for the
copy of $D_4$. As in the previous dimension, the separation between
adjacent layers will be $1$.

The $E_6$ lattice, which is the Bravais lattice with the highest
packing density, is given by the following coloring:
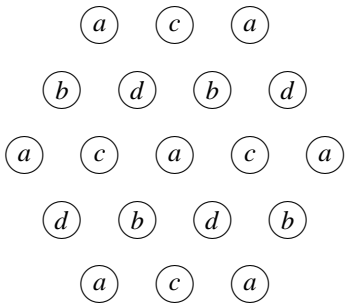
\begin{center}
\begin{picture}(4.25,4.052)(-0.25,0.128)
\put(0,2.1){\circle{0.5}}
\put(1,2.1){\circle{0.5}}
\put(2,2.1){\circle{0.5}}
\put(3,2.1){\circle{0.5}}
\put(4,2.1){\circle{0.5}}

\put(-0.5,1.6){\makebox(1,1){$a$}}
\put(0.5,1.6){\makebox(1,1){$c$}}
\put(1.5,1.6){\makebox(1,1){$a$}}
\put(2.5,1.6){\makebox(1,1){$c$}}
\put(3.5,1.6){\makebox(1,1){$a$}}

\put(1,3.832){\circle{0.5}}
\put(2,3.832){\circle{0.5}}
\put(3,3.832){\circle{0.5}}

\put(0.5,3.332){\makebox(1,1){$a$}}
\put(1.5,3.332){\makebox(1,1){$c$}}
\put(2.5,3.332){\makebox(1,1){$a$}}

\put(1,0.378){\circle{0.5}}
\put(2,0.378){\circle{0.5}}
\put(3,0.378){\circle{0.5}}

\put(0.5,-0.122){\makebox(1,1){$a$}}
\put(1.5,-0.122){\makebox(1,1){$c$}}
\put(2.5,-0.122){\makebox(1,1){$a$}}

\put(0.5,2.966){\circle{0.5}}
\put(1.5,2.966){\circle{0.5}}
\put(2.5,2.966){\circle{0.5}}
\put(3.5,2.966){\circle{0.5}}

\put(0,2.466){\makebox(1,1){$b$}}
\put(1,2.466){\makebox(1,1){$d$}}
\put(2,2.466){\makebox(1,1){$b$}}
\put(3,2.466){\makebox(1,1){$d$}}

\put(0.5,1.234){\circle{0.5}}
\put(1.5,1.234){\circle{0.5}}
\put(2.5,1.234){\circle{0.5}}
\put(3.5,1.234){\circle{0.5}}

\put(0,0.734){\makebox(1,1){$d$}}
\put(1,0.734){\makebox(1,1){$b$}}
\put(2,0.734){\makebox(1,1){$d$}}
\put(3,0.734){\makebox(1,1){$b$}}
\end{picture}
\end{center}
The theta series of $E_6$ is $1+72q^2 + 270q^4 + 936q^6 + 2160q^8 +
\cdots$. As shown by Conway and Sloane, there are three other uniform
packings, corresponding to the following possibilities for the six
neighbors surrounding a central $a$:
\begin{align*}
\Lambda_6^2 &\ :\  bcbdcd, \\
\Lambda_6^3 &\ :\  bcbcbc, \\
\Lambda_6^4 &\ :\  bcbcbd.
\end{align*}
In contrast to the five-dimensional case, the Bravais lattice $E_6$ is
in fact optimal among all tight lattices in the $q \to 0$ limit.

\begin{theorem}
Under Postulates 2, 4, and 6 of Ref.\ \cite{CS}, the Bravais lattice
$E_6$ has the lowest energy among all the tight six-dimensional
lattices, in the $q \to 0$ limit.
\end{theorem}

\begin{proof}
Let $\Lambda$ be a tight packing formed as above by four-coloring the
triangular lattice. Let us assume that the central sphere is colored
$a$. The squared distances in the $A_2$ lattice are $1,3,4,\dots$, so
neighbors at squared distance $2$ in $\Lambda$ can come only from the
central layer and its six adjacent layers. The number of these vectors
is $24 + 6 \cdot 8 = 72$, which is in accordance with the theta
function of $E_6$. The next possible squared distance is $3$. Note that
this distance does not occur in $E_6$, since in the coloring above,
there are no two spheres at squared distance $3$ which have the same
color. But in fact, the coloring above is the only coloring with this
property (up to a permutation of the colors $a$, $b$, $c$, $d$, of
course, but that is irrelevant because of the symmetries of $D_4$). To
see this, start with the central sphere colored $a$, and notice that
the six spheres around it must be colored $bcdbcd$ (or $bdcbdc$) to
avoid two spheres of the same color being $\sqrt{3}$ units apart. One
can then apply the argument to the six spheres centered around one of
these six neighbors and proceed outward, to arrive at a unique packing:
namely, the one above. This shows that $E_6$ is indeed the best for
energy in the $q \to 0$ limit, among all tight lattices.
\end{proof}

One can also determine the worst tight packing.

\begin{theorem}
Under Postulates 2, 4, and 6 of Ref.\ \cite{CS}, the lattice
$\Lambda_6^3$ has the highest energy among all the tight
six-dimensional lattices, in the $q \to 0$ limit.
\end{theorem}

We omit the details of the proof.  However, the calculation amounts to
showing that the $\Lambda_6^3$ coloring maximizes the number of
identically colored spheres at squared distance $3$ in $A_2$.  In the
following picture of the coloring, the six bold circles are at squared
distance $3$ from the central circle:

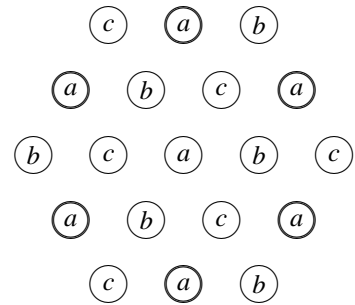
\begin{center}
\begin{picture}(4.25,4.052)(-0.25,0.128)
\put(1,3.832){\circle{0.5}}
\put(2,3.832){\circle{0.5}}
\put(2,3.832){\circle{0.45}}
\put(3,3.832){\circle{0.5}}

\put(0.5,3.332){\makebox(1,1){$c$}}
\put(1.5,3.332){\makebox(1,1){$a$}}
\put(2.5,3.332){\makebox(1,1){$b$}}

\put(0.5,2.966){\circle{0.5}}
\put(0.5,2.966){\circle{0.45}}
\put(1.5,2.966){\circle{0.5}}
\put(2.5,2.966){\circle{0.5}}
\put(3.5,2.966){\circle{0.5}}
\put(3.5,2.966){\circle{0.45}}

\put(0,2.466){\makebox(1,1){$a$}}
\put(1,2.466){\makebox(1,1){$b$}}
\put(2,2.466){\makebox(1,1){$c$}}
\put(3,2.466){\makebox(1,1){$a$}}

\put(0,2.1){\circle{0.5}}
\put(1,2.1){\circle{0.5}}
\put(2,2.1){\circle{0.5}}
\put(3,2.1){\circle{0.5}}
\put(4,2.1){\circle{0.5}}

\put(-0.5,1.6){\makebox(1,1){$b$}}
\put(0.5,1.6){\makebox(1,1){$c$}}
\put(1.5,1.6){\makebox(1,1){$a$}}
\put(2.5,1.6){\makebox(1,1){$b$}}
\put(3.5,1.6){\makebox(1,1){$c$}}

\put(0.5,1.234){\circle{0.5}}
\put(0.5,1.234){\circle{0.45}}
\put(1.5,1.234){\circle{0.5}}
\put(2.5,1.234){\circle{0.5}}
\put(3.5,1.234){\circle{0.5}}
\put(3.5,1.234){\circle{0.45}}

\put(0,0.734){\makebox(1,1){$a$}}
\put(1,0.734){\makebox(1,1){$b$}}
\put(2,0.734){\makebox(1,1){$c$}}
\put(3,0.734){\makebox(1,1){$a$}}

\put(1,0.378){\circle{0.5}}
\put(2,0.378){\circle{0.5}}
\put(2,0.378){\circle{0.45}}
\put(3,0.378){\circle{0.5}}

\put(0.5,-0.122){\makebox(1,1){$c$}}
\put(1.5,-0.122){\makebox(1,1){$a$}}
\put(2.5,-0.122){\makebox(1,1){$b$}}
\end{picture}
\end{center}

\section{Dimension $7$}

Finally, in dimension $7$ the optimal Bravais lattice $E_7$ is neither
the worst nor the best for energy among tight packings in the low
particle-density limit. According to Ref.\ \cite{CS}, each tight
packing in $\R^7$ fibers over $D_4$, and the four-dimensional layers
are arranged using a tight packing in $\R^3$ with adjacent $D_4$ layers
separated by $1$. To specify a four-coloring of the three-dimensional
packing, we need only specify it on a single triangular layer, since
each such layer determines the colors on both adjacent layers and hence
on every layer.

We cannot use an arbitrary four-coloring of the triangular layer, since
some colorings do not extend consistently to the other layers.  Conway
and Sloane showed that the condition for extending consistently is that
the coloring should have ``period 2'' in the following sense: the
packing should decompose into parallel strings of adjacent spheres, so
that in each string the colors alternate between two possibilities. For
example, the $E_6$ coloring shown in the previous section has this
property (the strings lie along horizontal lines), while the
$\Lambda_6^3$ coloring does not.

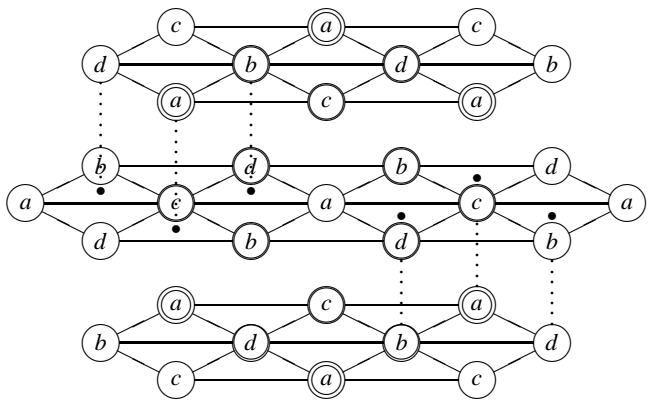
\begin{figure}
\begin{center}
\begin{picture}(8.5,5.2)(-0.25,-0.6)


\put(0,2){\circle{0.5}} \put(-0.5,1.5){\makebox(1,1){$a$}}
\put(2,2){\circle{0.5}} \put(1.5,1.5){\makebox(1,1){$c$}}
\put(4,2){\circle{0.5}} \put(3.5,1.5){\makebox(1,1){$a$}}
\put(6,2){\circle{0.5}} \put(5.5,1.5){\makebox(1,1){$c$}}
\put(8,2){\circle{0.5}} \put(7.5,1.5){\makebox(1,1){$a$}}

\put(1,2.5){\circle{0.5}} \put(0.5,2){\makebox(1,1){$b$}}
\put(3,2.5){\circle{0.5}} \put(2.5,2){\makebox(1,1){$d$}}
\put(5,2.5){\circle{0.5}} \put(4.5,2){\makebox(1,1){$b$}}
\put(7,2.5){\circle{0.5}} \put(6.5,2){\makebox(1,1){$d$}}

\put(1,1.5){\circle{0.5}} \put(0.5,1){\makebox(1,1){$d$}}
\put(3,1.5){\circle{0.5}} \put(2.5,1){\makebox(1,1){$b$}}
\put(5,1.5){\circle{0.5}} \put(4.5,1){\makebox(1,1){$d$}}
\put(7,1.5){\circle{0.5}} \put(6.5,1){\makebox(1,1){$b$}}

\put(0.224,2.112){\line(2,1){0.553}}
\put(2.224,2.112){\line(2,1){0.553}}
\put(4.224,2.112){\line(2,1){0.553}}
\put(6.224,2.112){\line(2,1){0.553}}

\put(1.224,1.612){\line(2,1){0.553}}
\put(3.224,1.612){\line(2,1){0.553}}
\put(5.224,1.612){\line(2,1){0.553}}
\put(7.224,1.612){\line(2,1){0.553}}

\put(0.224,1.888){\line(2,-1){0.553}}
\put(2.224,1.888){\line(2,-1){0.553}}
\put(4.224,1.888){\line(2,-1){0.553}}
\put(6.224,1.888){\line(2,-1){0.553}}

\put(1.224,2.388){\line(2,-1){0.553}}
\put(3.224,2.388){\line(2,-1){0.553}}
\put(5.224,2.388){\line(2,-1){0.553}}
\put(7.224,2.388){\line(2,-1){0.553}}

\put(0.25,2){\line(1,0){1.5}} \put(2.25,2){\line(1,0){1.5}}
\put(4.25,2){\line(1,0){1.5}} \put(6.25,2){\line(1,0){1.5}}

\put(1.25,2.5){\line(1,0){1.5}} \put(3.25,2.5){\line(1,0){1.5}}
\put(5.25,2.5){\line(1,0){1.5}}

\put(1.25,1.5){\line(1,0){1.5}} \put(3.25,1.5){\line(1,0){1.5}}
\put(5.25,1.5){\line(1,0){1.5}}


\put(1,3.85){\circle{0.5}} \put(0.5,3.35){\makebox(1,1){$d$}}
\put(3,3.85){\circle{0.5}} \put(2.5,3.35){\makebox(1,1){$b$}}
\put(5,3.85){\circle{0.5}} \put(4.5,3.35){\makebox(1,1){$d$}}
\put(7,3.85){\circle{0.5}} \put(6.5,3.35){\makebox(1,1){$b$}}

\put(2,3.35){\circle{0.5}} \put(2,3.35){\circle{0.4}}
\put(1.5,2.85){\makebox(1,1){${a}$}} \put(4,3.35){\circle{0.5}}
\put(3.5,2.85){\makebox(1,1){$c$}} \put(6,3.35){\circle{0.5}}
\put(6,3.35){\circle{0.4}} \put(5.5,2.85){\makebox(1,1){${a}$}}

\put(2,4.35){\circle{0.5}} \put(1.5,3.85){\makebox(1,1){$c$}}
\put(4,4.35){\circle{0.5}} \put(4,4.35){\circle{0.4}}
\put(3.5,3.85){\makebox(1,1){${a}$}} \put(6,4.35){\circle{0.5}}
\put(5.5,3.85){\makebox(1,1){$c$}}

\put(1.25,3.85){\line(1,0){1.5}} \put(3.25,3.85){\line(1,0){1.5}}
\put(5.25,3.85){\line(1,0){1.5}}

\put(2.25,3.35){\line(1,0){1.5}} \put(4.25,3.35){\line(1,0){1.5}}

\put(2.224,3.462){\line(2,1){0.553}}
\put(4.224,3.462){\line(2,1){0.553}}
\put(6.224,3.462){\line(2,1){0.553}}

\put(1.224,3.738){\line(2,-1){0.553}}
\put(3.224,3.738){\line(2,-1){0.553}}
\put(5.224,3.738){\line(2,-1){0.553}}

\put(1.224,3.962){\line(2,1){0.553}}
\put(3.224,3.962){\line(2,1){0.553}}
\put(5.224,3.962){\line(2,1){0.553}}

\put(2.224,4.238){\line(2,-1){0.553}}
\put(4.224,4.238){\line(2,-1){0.553}}
\put(6.224,4.238){\line(2,-1){0.553}}

\put(2.25,4.35){\line(1,0){1.5}} \put(4.25,4.35){\line(1,0){1.5}}


\put(1,0.15){\circle{0.5}} \put(0.5,-0.35){\makebox(1,1){$b$}}
\put(3,0.15){\circle{0.5}} \put(2.5,-0.35){\makebox(1,1){$d$}}
\put(5,0.15){\circle{0.5}} \put(4.5,-0.35){\makebox(1,1){$b$}}
\put(7,0.15){\circle{0.5}} \put(6.5,-0.35){\makebox(1,1){$d$}}

\put(2,0.65){\circle{0.5}} \put(2,0.65){\circle{0.4}}
\put(1.5,0.15){\makebox(1,1){${a}$}} \put(4,0.65){\circle{0.5}}
\put(3.5,0.15){\makebox(1,1){$c$}} \put(6,0.65){\circle{0.5}}
\put(6,0.65){\circle{0.4}} \put(5.5,0.15){\makebox(1,1){${a}$}}

\put(2,-0.35){\circle{0.5}} \put(1.5,-0.85){\makebox(1,1){$c$}}
\put(4,-0.35){\circle{0.5}} \put(4,-0.35){\circle{0.4}}
\put(3.5,-0.85){\makebox(1,1){${a}$}} \put(6,-0.35){\circle{0.5}}
\put(5.5,-0.85){\makebox(1,1){$c$}}

\put(1.25,0.15){\line(1,0){1.5}} \put(3.25,0.15){\line(1,0){1.5}}
\put(5.25,0.15){\line(1,0){1.5}}

\put(2.25,0.65){\line(1,0){1.5}} \put(4.25,0.65){\line(1,0){1.5}}

\put(1.224,0.262){\line(2,1){0.553}}
\put(3.224,0.262){\line(2,1){0.553}}
\put(5.224,0.262){\line(2,1){0.553}}

\put(2.224,0.538){\line(2,-1){0.553}}
\put(4.224,0.538){\line(2,-1){0.553}}
\put(6.224,0.538){\line(2,-1){0.553}}

\put(2.25,-0.35){\line(1,0){1.5}} \put(4.25,-0.35){\line(1,0){1.5}}

\put(2.224,-0.238){\line(2,1){0.553}}
\put(4.224,-0.238){\line(2,1){0.553}}
\put(6.224,-0.238){\line(2,1){0.553}}

\put(1.224,0.038){\line(2,-1){0.553}}
\put(3.224,0.038){\line(2,-1){0.553}}
\put(5.224,0.038){\line(2,-1){0.553}}


\put(1,2.166){\circle*{0.1}} \put(2,1.666){\circle*{0.1}}
\put(3,2.166){\circle*{0.1}}

\multiput(1,3.6)(0,-0.13854){10}{\circle*{0.05}}
\multiput(3,3.6)(0,-0.13854){10}{\circle*{0.05}}
\multiput(2,3.1)(0,-0.13854){10}{\circle*{0.05}}


\put(5,1.833){\circle*{0.1}} \put(6,2.333){\circle*{0.1}}
\put(7,1.833){\circle*{0.1}}

\multiput(5,0.4)(0,0.13854){7}{\circle*{0.05}}
\multiput(7,0.4)(0,0.13854){7}{\circle*{0.05}}
\multiput(6,0.9)(0,0.13854){7}{\circle*{0.05}}


\put(2,2){\circle{0.45}} \put(6,2){\circle{0.45}}
\put(3,2.5){\circle{0.45}} \put(5,2.5){\circle{0.45}}
\put(3,1.5){\circle{0.45}} \put(5,1.5){\circle{0.45}}

\put(3,3.85){\circle{0.45}} \put(4,3.35){\circle{0.45}}
\put(5,3.85){\circle{0.45}} \put(3,0.15){\circle{0.45}}
\put(4,0.65){\circle{0.45}} \put(5,0.15){\circle{0.45}}
\end{picture}
\end{center}
\caption{The four-coloring of the fcc lattice used to obtain $E_7$.}
\label{fig:e7}
\end{figure}

To obtain $E_7$, we use the face-centered cubic as our tight packing in
$\R^3$, and we use the same coloring of a triangular layer as was used
to construct $E_6$.  We get the picture in Fig.~\ref{fig:e7}, which
shows three triangular layers of the fcc lattice surrounding a central
ball colored $a$ (the dotted lines show how the layers are aligned, and
the different styles of circles are for reference in the argument
below).

The theta series of $E_7$ is $1 + 126q^2 + 756q^4 +  \cdots$, and we
can see the first nontrivial term as follows.  A point in the $D_4$
layer corresponding to the central circle colored $a$ above has $24$
neighbors at squared distance $2$ in the same $D_4$ layer, $12\cdot 8$
in neighboring $D_4$ layers ($8$ each from the $12$ neighbors in the
face-centered cubic, which have bold circles in Fig.~\ref{fig:e7}), and
$6$ from non-neighboring $D_4$ layers ($1$ each from the $6$ points in
the face-centered cubic at squared distance $2$, which are shown with
two nested circles in Fig.~\ref{fig:e7} and are each colored $a$).

\begin{figure}
\begin{center}
\begin{picture}(8.5,5.2)(-0.25,-0.6)


\put(0,2){\circle{0.5}} \put(-0.5,1.5){\makebox(1,1){$a$}}
\put(2,2){\circle{0.5}} \put(1.5,1.5){\makebox(1,1){$c$}}
\put(4,2){\circle{0.5}} \put(3.5,1.5){\makebox(1,1){$a$}}
\put(6,2){\circle{0.5}} \put(5.5,1.5){\makebox(1,1){$c$}}
\put(8,2){\circle{0.5}} \put(7.5,1.5){\makebox(1,1){$a$}}

\put(1,2.5){\circle{0.5}} \put(0.5,2){\makebox(1,1){$b$}}
\put(3,2.5){\circle{0.5}} \put(2.5,2){\makebox(1,1){$d$}}
\put(5,2.5){\circle{0.5}} \put(4.5,2){\makebox(1,1){$b$}}
\put(7,2.5){\circle{0.5}} \put(6.5,2){\makebox(1,1){$d$}}

\put(1,1.5){\circle{0.5}} \put(0.5,1){\makebox(1,1){$b$}}
\put(3,1.5){\circle{0.5}} \put(2.5,1){\makebox(1,1){$d$}}
\put(5,1.5){\circle{0.5}} \put(4.5,1){\makebox(1,1){$b$}}
\put(7,1.5){\circle{0.5}} \put(6.5,1){\makebox(1,1){$d$}}

\put(0.224,2.112){\line(2,1){0.553}}
\put(2.224,2.112){\line(2,1){0.553}}
\put(4.224,2.112){\line(2,1){0.553}}
\put(6.224,2.112){\line(2,1){0.553}}

\put(1.224,1.612){\line(2,1){0.553}}
\put(3.224,1.612){\line(2,1){0.553}}
\put(5.224,1.612){\line(2,1){0.553}}
\put(7.224,1.612){\line(2,1){0.553}}

\put(0.224,1.888){\line(2,-1){0.553}}
\put(2.224,1.888){\line(2,-1){0.553}}
\put(4.224,1.888){\line(2,-1){0.553}}
\put(6.224,1.888){\line(2,-1){0.553}}

\put(1.224,2.388){\line(2,-1){0.553}}
\put(3.224,2.388){\line(2,-1){0.553}}
\put(5.224,2.388){\line(2,-1){0.553}}
\put(7.224,2.388){\line(2,-1){0.553}}

\put(0.25,2){\line(1,0){1.5}} \put(2.25,2){\line(1,0){1.5}}
\put(4.25,2){\line(1,0){1.5}} \put(6.25,2){\line(1,0){1.5}}

\put(1.25,2.5){\line(1,0){1.5}} \put(3.25,2.5){\line(1,0){1.5}}
\put(5.25,2.5){\line(1,0){1.5}}

\put(1.25,1.5){\line(1,0){1.5}} \put(3.25,1.5){\line(1,0){1.5}}
\put(5.25,1.5){\line(1,0){1.5}}


\put(1,3.85){\circle{0.5}} \put(0.5,3.35){\makebox(1,1){$d$}}
\put(3,3.85){\circle{0.5}} \put(2.5,3.35){\makebox(1,1){$b$}}
\put(5,3.85){\circle{0.5}} \put(4.5,3.35){\makebox(1,1){$d$}}
\put(7,3.85){\circle{0.5}} \put(6.5,3.35){\makebox(1,1){$b$}}

\put(2,3.35){\circle{0.5}} \put(2,3.35){\circle{0.4}}
\put(1.5,2.85){\makebox(1,1){${a}$}} \put(4,3.35){\circle{0.5}}
\put(3.5,2.85){\makebox(1,1){$c$}} \put(6,3.35){\circle{0.5}}
\put(6,3.35){\circle{0.4}} \put(5.5,2.85){\makebox(1,1){${a}$}}

\put(2,4.35){\circle{0.5}} \put(1.5,3.85){\makebox(1,1){$a$}}
\put(4,4.35){\circle{0.5}} \put(3.5,3.85){\makebox(1,1){${c}$}}
\put(6,4.35){\circle{0.5}} \put(5.5,3.85){\makebox(1,1){$a$}}

\put(1.25,3.85){\line(1,0){1.5}} \put(3.25,3.85){\line(1,0){1.5}}
\put(5.25,3.85){\line(1,0){1.5}}

\put(2.25,3.35){\line(1,0){1.5}} \put(4.25,3.35){\line(1,0){1.5}}

\put(2.224,3.462){\line(2,1){0.553}}
\put(4.224,3.462){\line(2,1){0.553}}
\put(6.224,3.462){\line(2,1){0.553}}

\put(1.224,3.738){\line(2,-1){0.553}}
\put(3.224,3.738){\line(2,-1){0.553}}
\put(5.224,3.738){\line(2,-1){0.553}}

\put(1.224,3.962){\line(2,1){0.553}}
\put(3.224,3.962){\line(2,1){0.553}}
\put(5.224,3.962){\line(2,1){0.553}}

\put(2.224,4.238){\line(2,-1){0.553}}
\put(4.224,4.238){\line(2,-1){0.553}}
\put(6.224,4.238){\line(2,-1){0.553}}

\put(2.25,4.35){\line(1,0){1.5}} \put(4.25,4.35){\line(1,0){1.5}}


\put(1,0.15){\circle{0.5}} \put(0.5,-0.35){\makebox(1,1){$d$}}
\put(3,0.15){\circle{0.5}} \put(2.5,-0.35){\makebox(1,1){$b$}}
\put(5,0.15){\circle{0.5}} \put(4.5,-0.35){\makebox(1,1){$d$}}
\put(7,0.15){\circle{0.5}} \put(6.5,-0.35){\makebox(1,1){$b$}}

\put(2,0.65){\circle{0.5}} \put(2,0.65){\circle{0.4}}
\put(1.5,0.15){\makebox(1,1){${a}$}} \put(4,0.65){\circle{0.5}}
\put(3.5,0.15){\makebox(1,1){$c$}} \put(6,0.65){\circle{0.5}}
\put(6,0.65){\circle{0.4}} \put(5.5,0.15){\makebox(1,1){${a}$}}

\put(2,-0.35){\circle{0.5}} \put(1.5,-0.85){\makebox(1,1){$a$}}
\put(4,-0.35){\circle{0.5}} \put(3.5,-0.85){\makebox(1,1){${c}$}}
\put(6,-0.35){\circle{0.5}} \put(5.5,-0.85){\makebox(1,1){$a$}}

\put(1.25,0.15){\line(1,0){1.5}} \put(3.25,0.15){\line(1,0){1.5}}
\put(5.25,0.15){\line(1,0){1.5}}

\put(2.25,0.65){\line(1,0){1.5}} \put(4.25,0.65){\line(1,0){1.5}}

\put(1.224,0.262){\line(2,1){0.553}}
\put(3.224,0.262){\line(2,1){0.553}}
\put(5.224,0.262){\line(2,1){0.553}}

\put(2.224,0.538){\line(2,-1){0.553}}
\put(4.224,0.538){\line(2,-1){0.553}}
\put(6.224,0.538){\line(2,-1){0.553}}

\put(2.25,-0.35){\line(1,0){1.5}} \put(4.25,-0.35){\line(1,0){1.5}}

\put(2.224,-0.238){\line(2,1){0.553}}
\put(4.224,-0.238){\line(2,1){0.553}}
\put(6.224,-0.238){\line(2,1){0.553}}

\put(1.224,0.038){\line(2,-1){0.553}}
\put(3.224,0.038){\line(2,-1){0.553}}
\put(5.224,0.038){\line(2,-1){0.553}}


\put(1,2.166){\circle*{0.1}} \put(2,1.666){\circle*{0.1}}
\put(3,2.166){\circle*{0.1}}

\multiput(1,3.6)(0,-0.13854){10}{\circle*{0.05}}
\multiput(3,3.6)(0,-0.13854){10}{\circle*{0.05}}
\multiput(2,3.1)(0,-0.13854){10}{\circle*{0.05}}


\put(5,1.833){\circle*{0.1}} \put(6,2.333){\circle*{0.1}}
\put(7,1.833){\circle*{0.1}}

\multiput(5,0.4)(0,0.13854){7}{\circle*{0.05}}
\multiput(7,0.4)(0,0.13854){7}{\circle*{0.05}}
\multiput(6,0.9)(0,0.13854){7}{\circle*{0.05}}


\put(2,2){\circle{0.45}} \put(6,2){\circle{0.45}}
\put(3,2.5){\circle{0.45}} \put(5,2.5){\circle{0.45}}
\put(3,1.5){\circle{0.45}} \put(5,1.5){\circle{0.45}}

\put(3,3.85){\circle{0.45}} \put(4,3.35){\circle{0.45}}
\put(5,3.85){\circle{0.45}} \put(3,0.15){\circle{0.45}}
\put(4,0.65){\circle{0.45}} \put(5,0.15){\circle{0.45}}
\end{picture}
\end{center}
\caption{The four-coloring of the fcc lattice used to obtained
$\Lambda_7^3$.} \label{fig:Lambda73}
\end{figure}

To improve upon $E_7$ in the $q \to 0$ limit, we can use the
$\Lambda_6^2$ coloring of a triangular layer; the resulting tight
packing is called $\Lambda_7^3$. One can see from
Fig.~\ref{fig:Lambda73} that among the final six points in the
calculation above, only four share the same color as the central point.
Thus, the theta series of $\Lambda_7^3$ begins $1+ 124q^2 + \cdots$,
which is an improvement over the $E_7$ lattice, and the next squared
distance is $4$.

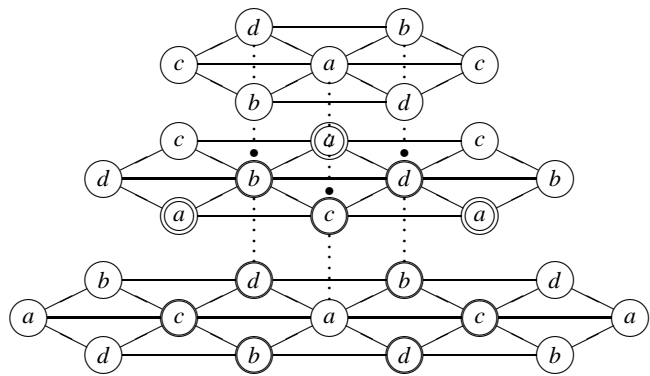
\begin{figure}
\begin{center}
\begin{picture}(8.5,4.867)(-0.25,1.25)


\put(0,2){\circle{0.5}} \put(-0.5,1.5){\makebox(1,1){$a$}}
\put(2,2){\circle{0.5}} \put(1.5,1.5){\makebox(1,1){$c$}}
\put(4,2){\circle{0.5}} \put(3.5,1.5){\makebox(1,1){$a$}}
\put(6,2){\circle{0.5}} \put(5.5,1.5){\makebox(1,1){$c$}}
\put(8,2){\circle{0.5}} \put(7.5,1.5){\makebox(1,1){$a$}}

\put(1,2.5){\circle{0.5}} \put(0.5,2){\makebox(1,1){$b$}}
\put(3,2.5){\circle{0.5}} \put(2.5,2){\makebox(1,1){$d$}}
\put(5,2.5){\circle{0.5}} \put(4.5,2){\makebox(1,1){$b$}}
\put(7,2.5){\circle{0.5}} \put(6.5,2){\makebox(1,1){$d$}}

\put(1,1.5){\circle{0.5}} \put(0.5,1){\makebox(1,1){$d$}}
\put(3,1.5){\circle{0.5}} \put(2.5,1){\makebox(1,1){$b$}}
\put(5,1.5){\circle{0.5}} \put(4.5,1){\makebox(1,1){$d$}}
\put(7,1.5){\circle{0.5}} \put(6.5,1){\makebox(1,1){$b$}}

\put(0.224,2.112){\line(2,1){0.553}}
\put(2.224,2.112){\line(2,1){0.553}}
\put(4.224,2.112){\line(2,1){0.553}}
\put(6.224,2.112){\line(2,1){0.553}}

\put(1.224,1.612){\line(2,1){0.553}}
\put(3.224,1.612){\line(2,1){0.553}}
\put(5.224,1.612){\line(2,1){0.553}}
\put(7.224,1.612){\line(2,1){0.553}}

\put(0.224,1.888){\line(2,-1){0.553}}
\put(2.224,1.888){\line(2,-1){0.553}}
\put(4.224,1.888){\line(2,-1){0.553}}
\put(6.224,1.888){\line(2,-1){0.553}}

\put(1.224,2.388){\line(2,-1){0.553}}
\put(3.224,2.388){\line(2,-1){0.553}}
\put(5.224,2.388){\line(2,-1){0.553}}
\put(7.224,2.388){\line(2,-1){0.553}}

\put(0.25,2){\line(1,0){1.5}} \put(2.25,2){\line(1,0){1.5}}
\put(4.25,2){\line(1,0){1.5}} \put(6.25,2){\line(1,0){1.5}}

\put(1.25,2.5){\line(1,0){1.5}} \put(3.25,2.5){\line(1,0){1.5}}
\put(5.25,2.5){\line(1,0){1.5}}

\put(1.25,1.5){\line(1,0){1.5}} \put(3.25,1.5){\line(1,0){1.5}}
\put(5.25,1.5){\line(1,0){1.5}}


\put(1,3.85){\circle{0.5}} \put(0.5,3.35){\makebox(1,1){$d$}}
\put(3,3.85){\circle{0.5}} \put(2.5,3.35){\makebox(1,1){$b$}}
\put(5,3.85){\circle{0.5}} \put(4.5,3.35){\makebox(1,1){$d$}}
\put(7,3.85){\circle{0.5}} \put(6.5,3.35){\makebox(1,1){$b$}}

\put(2,3.35){\circle{0.5}} \put(2,3.35){\circle{0.4}}
\put(1.5,2.85){\makebox(1,1){${a}$}} \put(4,3.35){\circle{0.5}}
\put(3.5,2.85){\makebox(1,1){$c$}} \put(6,3.35){\circle{0.5}}
\put(6,3.35){\circle{0.4}} \put(5.5,2.85){\makebox(1,1){${a}$}}

\put(2,4.35){\circle{0.5}} \put(1.5,3.85){\makebox(1,1){$c$}}
\put(4,4.35){\circle{0.5}} \put(4,4.35){\circle{0.4}}
\put(3.5,3.85){\makebox(1,1){${a}$}} \put(6,4.35){\circle{0.5}}
\put(5.5,3.85){\makebox(1,1){$c$}}

\put(1.25,3.85){\line(1,0){1.5}} \put(3.25,3.85){\line(1,0){1.5}}
\put(5.25,3.85){\line(1,0){1.5}}

\put(2.25,3.35){\line(1,0){1.5}} \put(4.25,3.35){\line(1,0){1.5}}

\put(2.224,3.462){\line(2,1){0.553}}
\put(4.224,3.462){\line(2,1){0.553}}
\put(6.224,3.462){\line(2,1){0.553}}

\put(1.224,3.738){\line(2,-1){0.553}}
\put(3.224,3.738){\line(2,-1){0.553}}
\put(5.224,3.738){\line(2,-1){0.553}}

\put(1.224,3.962){\line(2,1){0.553}}
\put(3.224,3.962){\line(2,1){0.553}}
\put(5.224,3.962){\line(2,1){0.553}}

\put(2.224,4.238){\line(2,-1){0.553}}
\put(4.224,4.238){\line(2,-1){0.553}}
\put(6.224,4.238){\line(2,-1){0.553}}

\put(2.25,4.35){\line(1,0){1.5}} \put(4.25,4.35){\line(1,0){1.5}}


\put(2,5.367){\circle{0.5}} \put(1.5,4.867){\makebox(1,1){$c$}}
\put(4,5.367){\circle{0.5}} \put(3.5,4.867){\makebox(1,1){$a$}}
\put(6,5.367){\circle{0.5}} \put(5.5,4.867){\makebox(1,1){$c$}}

\put(3,5.867){\circle{0.5}} \put(2.5,5.367){\makebox(1,1){$d$}}
\put(5,5.867){\circle{0.5}} \put(4.5,5.367){\makebox(1,1){$b$}}

\put(3,4.867){\circle{0.5}} \put(2.5,4.367){\makebox(1,1){$b$}}
\put(5,4.867){\circle{0.5}} \put(4.5,4.367){\makebox(1,1){$d$}}

\put(2.224,5.478){\line(2,1){0.553}}
\put(4.224,5.478){\line(2,1){0.553}}

\put(3.224,4.978){\line(2,1){0.553}}
\put(5.224,4.978){\line(2,1){0.553}}

\put(2.224,5.254){\line(2,-1){0.553}}
\put(4.224,5.254){\line(2,-1){0.553}}

\put(3.224,5.754){\line(2,-1){0.553}}
\put(5.224,5.754){\line(2,-1){0.553}}

\put(2.25,5.367){\line(1,0){1.5}} \put(4.25,5.367){\line(1,0){1.5}}

\put(3.25,5.867){\line(1,0){1.5}} \put(3.25,4.867){\line(1,0){1.5}}


\put(4,3.683){\circle*{0.1}} \put(3,4.183){\circle*{0.1}}
\put(5,4.183){\circle*{0.1}}

\multiput(4,2.25)(0,0.13854){7}{\circle*{0.05}}
\multiput(3,2.75)(0,0.13854){7}{\circle*{0.05}}
\multiput(5,2.75)(0,0.13854){7}{\circle*{0.05}}

\multiput(3,5.617)(0,-0.13854){4}{\circle*{0.05}}
\multiput(5,5.617)(0,-0.13854){4}{\circle*{0.05}}
\multiput(4,5.117)(0,-0.13854){10}{\circle*{0.05}}

\multiput(3,4.50868)(0,-0.13854){2}{\circle*{0.05}}
\multiput(5,4.50868)(0,-0.13854){2}{\circle*{0.05}}


\put(2,2){\circle{0.45}} \put(6,2){\circle{0.45}}
\put(3,2.5){\circle{0.45}} \put(5,2.5){\circle{0.45}}
\put(3,1.5){\circle{0.45}} \put(5,1.5){\circle{0.45}}

\put(3,3.85){\circle{0.45}} \put(4,3.35){\circle{0.45}}
\put(5,3.85){\circle{0.45}}
\end{picture}
\end{center}
\caption{The four-coloring of the hcp lattice used to obtained
$\Lambda_7^2$.} \label{fig:Lambda72}
\end{figure}

To construct a tight packing with higher energy than $E_7$ in the
low-density limit, we can use the hexagonal close packing in $\R^3$,
while using the same coloring on a triangular layer as for $E_7$
(namely, the one also used to construct $E_6$).  The resulting coloring
is shown in Fig.~\ref{fig:Lambda72}, and the packing is called
$\Lambda_7^2$. The large triangular layer at the bottom of the figure
plays the same role as the central layer in the previous figures. We
have not drawn the layers below it because the hcp packing is mirror
symmetric about each layer.

The theta series begins $1 + 126q^2 + \cdots$ for the same reason as
above, but the next term is $2q^{8/3}$, which occurs between
nonadjacent triangular layers.  Specifically, each point in the hcp
packing is at distance $\sqrt{8/3}$ (i.e., twice the height
$\sqrt{2/3}$ of a regular tetrahedron with edge length $1$) from two
points, which are two layers above and below it. The dotted lines in
Fig.~\ref{fig:Lambda72} connect such points. Because the corresponding
points always have the same color, the theta series of $\Lambda_7^2$
beings $1 + 126q^2 + 2q^{8/3} + \cdots$, and hence $\Lambda_7^2$ has
higher energy than $E_7$ in the $q \to 0$ limit.

\begin{figure}
\begin{center}
\begin{picture}(8.5,4.867)(-0.25,1.25)


\put(0,2){\circle{0.5}} \put(-0.5,1.5){\makebox(1,1){$a$}}
\put(2,2){\circle{0.5}} \put(1.5,1.5){\makebox(1,1){$c$}}
\put(4,2){\circle{0.5}} \put(3.5,1.5){\makebox(1,1){$a$}}
\put(6,2){\circle{0.5}} \put(5.5,1.5){\makebox(1,1){$c$}}
\put(8,2){\circle{0.5}} \put(7.5,1.5){\makebox(1,1){$a$}}

\put(1,2.5){\circle{0.5}} \put(0.5,2){\makebox(1,1){$b$}}
\put(3,2.5){\circle{0.5}} \put(2.5,2){\makebox(1,1){$d$}}
\put(5,2.5){\circle{0.5}} \put(4.5,2){\makebox(1,1){$b$}}
\put(7,2.5){\circle{0.5}} \put(6.5,2){\makebox(1,1){$d$}}

\put(1,1.5){\circle{0.5}} \put(0.5,1){\makebox(1,1){$b$}}
\put(3,1.5){\circle{0.5}} \put(2.5,1){\makebox(1,1){$d$}}
\put(5,1.5){\circle{0.5}} \put(4.5,1){\makebox(1,1){$b$}}
\put(7,1.5){\circle{0.5}} \put(6.5,1){\makebox(1,1){$d$}}

\put(0.224,2.112){\line(2,1){0.553}}
\put(2.224,2.112){\line(2,1){0.553}}
\put(4.224,2.112){\line(2,1){0.553}}
\put(6.224,2.112){\line(2,1){0.553}}

\put(1.224,1.612){\line(2,1){0.553}}
\put(3.224,1.612){\line(2,1){0.553}}
\put(5.224,1.612){\line(2,1){0.553}}
\put(7.224,1.612){\line(2,1){0.553}}

\put(0.224,1.888){\line(2,-1){0.553}}
\put(2.224,1.888){\line(2,-1){0.553}}
\put(4.224,1.888){\line(2,-1){0.553}}
\put(6.224,1.888){\line(2,-1){0.553}}

\put(1.224,2.388){\line(2,-1){0.553}}
\put(3.224,2.388){\line(2,-1){0.553}}
\put(5.224,2.388){\line(2,-1){0.553}}
\put(7.224,2.388){\line(2,-1){0.553}}

\put(0.25,2){\line(1,0){1.5}} \put(2.25,2){\line(1,0){1.5}}
\put(4.25,2){\line(1,0){1.5}} \put(6.25,2){\line(1,0){1.5}}

\put(1.25,2.5){\line(1,0){1.5}} \put(3.25,2.5){\line(1,0){1.5}}
\put(5.25,2.5){\line(1,0){1.5}}

\put(1.25,1.5){\line(1,0){1.5}} \put(3.25,1.5){\line(1,0){1.5}}
\put(5.25,1.5){\line(1,0){1.5}}


\put(1,3.85){\circle{0.5}} \put(0.5,3.35){\makebox(1,1){$d$}}
\put(3,3.85){\circle{0.5}} \put(2.5,3.35){\makebox(1,1){$b$}}
\put(5,3.85){\circle{0.5}} \put(4.5,3.35){\makebox(1,1){$d$}}
\put(7,3.85){\circle{0.5}} \put(6.5,3.35){\makebox(1,1){$b$}}

\put(2,3.35){\circle{0.5}} \put(2,3.35){\circle{0.4}}
\put(1.5,2.85){\makebox(1,1){${a}$}} \put(4,3.35){\circle{0.5}}
\put(3.5,2.85){\makebox(1,1){$c$}} \put(6,3.35){\circle{0.5}}
\put(6,3.35){\circle{0.4}} \put(5.5,2.85){\makebox(1,1){${a}$}}

\put(2,4.35){\circle{0.5}} \put(1.5,3.85){\makebox(1,1){$a$}}
\put(4,4.35){\circle{0.5}} \put(3.5,3.85){\makebox(1,1){${c}$}}
\put(6,4.35){\circle{0.5}} \put(5.5,3.85){\makebox(1,1){$a$}}

\put(1.25,3.85){\line(1,0){1.5}} \put(3.25,3.85){\line(1,0){1.5}}
\put(5.25,3.85){\line(1,0){1.5}}

\put(2.25,3.35){\line(1,0){1.5}} \put(4.25,3.35){\line(1,0){1.5}}

\put(2.224,3.462){\line(2,1){0.553}}
\put(4.224,3.462){\line(2,1){0.553}}
\put(6.224,3.462){\line(2,1){0.553}}

\put(1.224,3.738){\line(2,-1){0.553}}
\put(3.224,3.738){\line(2,-1){0.553}}
\put(5.224,3.738){\line(2,-1){0.553}}

\put(1.224,3.962){\line(2,1){0.553}}
\put(3.224,3.962){\line(2,1){0.553}}
\put(5.224,3.962){\line(2,1){0.553}}

\put(2.224,4.238){\line(2,-1){0.553}}
\put(4.224,4.238){\line(2,-1){0.553}}
\put(6.224,4.238){\line(2,-1){0.553}}

\put(2.25,4.35){\line(1,0){1.5}} \put(4.25,4.35){\line(1,0){1.5}}


\put(2,5.367){\circle{0.5}} \put(1.5,4.867){\makebox(1,1){$c$}}
\put(4,5.367){\circle{0.5}} \put(3.5,4.867){\makebox(1,1){$a$}}
\put(6,5.367){\circle{0.5}} \put(5.5,4.867){\makebox(1,1){$c$}}

\put(3,5.867){\circle{0.5}} \put(2.5,5.367){\makebox(1,1){$d$}}
\put(5,5.867){\circle{0.5}} \put(4.5,5.367){\makebox(1,1){$b$}}

\put(3,4.867){\circle{0.5}} \put(2.5,4.367){\makebox(1,1){$d$}}
\put(5,4.867){\circle{0.5}} \put(4.5,4.367){\makebox(1,1){$b$}}

\put(2.224,5.478){\line(2,1){0.553}}
\put(4.224,5.478){\line(2,1){0.553}}

\put(3.224,4.978){\line(2,1){0.553}}
\put(5.224,4.978){\line(2,1){0.553}}

\put(2.224,5.254){\line(2,-1){0.553}}
\put(4.224,5.254){\line(2,-1){0.553}}

\put(3.224,5.754){\line(2,-1){0.553}}
\put(5.224,5.754){\line(2,-1){0.553}}

\put(2.25,5.367){\line(1,0){1.5}} \put(4.25,5.367){\line(1,0){1.5}}

\put(3.25,5.867){\line(1,0){1.5}} \put(3.25,4.867){\line(1,0){1.5}}


\put(4,3.683){\circle*{0.1}} \put(3,4.183){\circle*{0.1}}
\put(5,4.183){\circle*{0.1}}

\multiput(4,2.25)(0,0.13854){7}{\circle*{0.05}}
\multiput(3,2.75)(0,0.13854){7}{\circle*{0.05}}
\multiput(5,2.75)(0,0.13854){7}{\circle*{0.05}}

\multiput(3,5.617)(0,-0.13854){4}{\circle*{0.05}}
\multiput(5,5.617)(0,-0.13854){4}{\circle*{0.05}}
\multiput(4,5.117)(0,-0.13854){10}{\circle*{0.05}}

\multiput(3,4.50868)(0,-0.13854){2}{\circle*{0.05}}
\multiput(5,4.50868)(0,-0.13854){2}{\circle*{0.05}}


\put(2,2){\circle{0.45}} \put(6,2){\circle{0.45}}
\put(3,2.5){\circle{0.45}} \put(5,2.5){\circle{0.45}}
\put(3,1.5){\circle{0.45}} \put(5,1.5){\circle{0.45}}

\put(3,3.85){\circle{0.45}} \put(4,3.35){\circle{0.45}}
\put(5,3.85){\circle{0.45}}
\end{picture}
\end{center}
\caption{The four-coloring of the hcp lattice used to obtained
$\Lambda_7^4$.} \label{fig:Lambda74}
\end{figure}
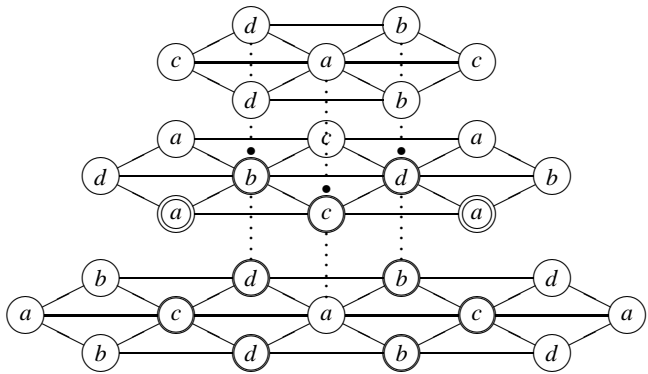

There is one further possibility worth analyzing, namely the coloring
of the hcp lattice shown in Fig.~\ref{fig:Lambda74} (which uses the
$\Lambda_6^2$ coloring on a triangular layer and leads to a tight
packing called $\Lambda_7^4$). Its theta series begins $1 + 124q^2 +
2q^{8/3} + \cdots$.

The four tight packings we have analyzed in this section are of course
not the only tight packings, but they are the only uniform ones. Their
local configurations cover enough possibilities to determine the
lowest- and highest-energy tight packings.  Specifically, there are
relatively few period $2$ colorings of a triangular layer. Observe the
large triangular layers in the figures: without loss of generality we
can assume that the middle horizontal row in the large triangular layer
is colored $acaca$ (by the period $2$ assumption), as is shown in each
figure. Then there are only two variables in the pictures. The first is
whether the adjacent two horizontal rows line up with $b$ above $b$ and
$d$ above $d$ (as in Figs.~\ref{fig:Lambda73} and~\ref{fig:Lambda74})
or whether they are staggered (as in the remaining two figures).  The
second variable is whether the triangular layers are themselves
staggered (as in the fcc lattice) or mirror-symmetric (as in the hcp
lattice).  If the pictures were to be enlarged, more of these choices
would arise, but within the scope of what has been drawn, there are
only these four possibilities. It follows that $\Lambda_7^3$ has the
best local configuration at each point, while $\Lambda_7^2$ has the
worst at each point.

\begin{theorem}
Under Postulates 2, 3, 4, and 7 of Ref.\ \cite{CS}, the lattice
$\Lambda_7^3$ has the lowest energy among all the tight
seven-dimensional lattices, in the $q \to 0$ limit, and $\Lambda_7^2$
has the highest.
\end{theorem}

\section{Higher dimensions}

In $\R^8$ there is a unique tight packing, namely the $E_8$ lattice,
which is almost certainly the ground state for the Gaussian core model.
Because of the uniqueness of $E_8$, the approach used in $\R^5$ and
$\R^7$  does not apply.

Above dimension $8$, the approach of Ref.\ \cite{CS} breaks down, and
tight packings no longer fiber nicely. Outside of a handful of
exceptional dimensions (certainly $24$ and perhaps $12$ or $16$), we
expect that the ground states of the Gaussian core model become quite
complicated.

\section{Conclusions and discussion}

We have shown that the ground states of the Gaussian core model can be
unexpectedly complex.  Specifically, in five and seven dimensions, the
ground states are not Bravais lattices, which contrasts with the more
familiar behavior in two or three dimensions.  This behavior is not
limited to the Gaussian core model.  The non-Bravais lattices studied
in this paper are in fact superior for a wide range of soft-core
models, including for example inverse power laws with high exponents.
(Note that inverse power laws are scale-free, so in that case our
results hold for all densities.)

These phenomena are characteristic of high dimensions, and they provide
support for the Torquato-Stillinger decorrelation principle.  As the
dimension increases, familiar symmetries become increasingly likely to
be broken. One noteworthy example is the kissing configurations in five
dimensions (i.e., the spherical configurations formed by the points of
tangency with adjacent spheres).  The $D_5$ lattice's kissing
configuration is highly symmetrical; in suitable coordinates it is
given by the vectors $(\pm 1 , \pm 1, 0, 0, 0)$ and all vectors
obtained by permuting the coordinates.  By contrast, the kissing
configuration of $\Lambda_5^2$ is far less symmetrical.  To form it,
replace the eight vectors that have a $1$ in the first coordinate with
the eight vectors $(1, \pm 1/2, \pm 1/2, \pm 1/2, \pm 1/2)$, where the
number of minus signs must be even.  This clearly breaks the symmetry,
and indeed the size of the symmetry group is reduced by a factor of
$10$ (from $3840$ to $384$). Nevertheless, $\Lambda_5^2$ has lower
energy than $D_5$, and its kissing configuration alone has lower energy
than that of $D_5$ as spherical configurations.  Symmetry simply does
not align with considerations of energy.

Because of the connections between high-dimensional sphere packing and
information theory, these issues shed light on coding theory.  Computer
scientists and engineers have learned through long experience that
efficient error-correcting codes should be chosen to be pseudo-random
(truly random would be even better, but it is generally not practical).
For example, MacKay \cite[p.~596]{Mac} summarizes his coding theory
advice as follows: ``The best solution to the communication problem is:
Combine a simple, pseudo-random code with a message-passing decoder.''
From a naive perspective, this situation is puzzling, since one might
expect that highly structured codes would offer the most scope for
powerful algorithms.  Instead, elaborate algebraic structure seems
incompatible with high-performance coding.  This is not purely a
geometric question, because of the role of algorithms, but it is
largely geometric, and the underlying geometry involves the same
decorrelation effect observed in physics.  This emphasizes the need for
a detailed theoretical understanding of high-dimensional packing and
related statistical mechanics models.

One natural area for further exploration would be non-Euclidean spaces.
Introducing curvature illuminates the problem of geometrical
frustration, in which ideal local configurations do not extend
consistently to global arrangements.  Specifically, curvature may
relieve (or introduce) frustration, and comparing results in different
curvatures clarifies the role of frustration.   See, for example, Ref.\
\cite{SM}.  Much work has been done in positively curved spaces such as
spheres, and Modes and Kamien \cite{MK1,MK2} have recently studied
hard-core models in negatively curved two-dimensional space.  It would
be intriguing to extend this work to higher dimensions.

Another area for future investigation is more sophisticated models than
the Gaussian core model.  For example, in the Ziherl-Kamien theory of
micellar crystals \cite{ZK1,ZK2}, area-minimizing effects (as in soap
froths) frustrate the close-packing one expects from a hard core.  It
would be interesting to study dimensional trends in such systems.

We conclude with a few specific open problems about the Gaussian core
model.

(1)  We have been able to address the low-density limit, but our
approach says nothing about the high-density limit.  Are Bravais
lattices optimal for the Gaussian core model at high density in low
dimensions, as Torquato and Stillinger \cite{TS2} conjectured?  We
suspect that Bravais lattices may again be suboptimal in as few as five
dimensions, but that is merely a guess.

(2)  In this paper, we were lucky to be able to construct improved
non-Bravais lattices essentially by careful modification of the Bravais
lattices (much as the hcp packing can be obtained by modifying the fcc
lattice). It is unlikely that this sort of modification will yield a
complete picture of the Gaussian core model's ground states at all
densities. In the absence of new geometrical ideas, it is natural to
turn to numerical simulations.  Unfortunately, simulations become
increasingly difficult as the dimension increases, because of the curse
of dimensionality (the number of particles required increases
exponentially as a function of dimension).  Can one develop an
efficient enough simulator to perform useful work in four, five, or
even six dimensions?  Skoge, Donev, Stillinger, and Torquato
\cite{SDST} have performed such simulations to compute jammed hard-core
packings, but that problem may be somewhat easier as there are no
long-range interactions.

(3)  Is the $D_4$ lattice universally optimal in $\R^4$?  (In other
words, is it the ground state of the Gaussian core model at every
density?) All available evidence suggests that the answer is yes,
except for one observation of Cohn, Conway, Elkies, and Kumar
\cite{CCEK}.  They show that the $D_4$ kissing configuration of $24$
points does not form a universally optimal spherical configuration, by
finding a competing family of configurations that occasionally beats
it. (By contrast, Cohn and Kumar \cite{CK} proved that the $E_8$
kissing configuration is universally optimal.) Unfortunately, the
spherical competitors do not seem to extend to Euclidean packings.
Because $D_4$ is such a symmetrical and beautiful structure, it would
be interesting to know more definitively whether it is universally
optimal.  Simulations could help resolve this issue.

\begin{acknowledgments}
We thank Salvatore Torquato and Frank Stillinger for helpful
discussions and the referees for their suggestions. A.K.\ was supported
in part by National Science Foundation Grant No.\ DMS-0757765.
\end{acknowledgments}



\vfill

\begin{thebibliography}{14}
\bibitem{S} F.~H.~Stillinger, J.\ Chem.\ Phys.\ \textbf{65}, 3968
    (1976).

\bibitem{FK} P.~J.~Flory and W.~R.~Krigbaum, J.\ Chem.\ Phys.\
    \textbf{18}, 1086 (1950).

\bibitem{LBHM} A.~A.~Louis, P.~G.~Bolhuis, J.~P.~Hansen, and
    E.~J.~Meijer, Phys.\ Rev.\ Lett.\ \textbf{85}, 2522 (2000).

\bibitem{LLWL} A.~Lang, C.~N.~Likos, M.~Watzlawek, and H.~L\"owen, J.\
    Phys.: Condens.\ Matter \textbf{12}, 5087 (2000).

\bibitem{PSG} S.~Prestipino, F.~Saija, and P.~V.~Giaquinta, Phys.\
    Rev.\ E \textbf{71}, 050102(R) (2005).

\bibitem{T} F.~Theil, Commun.\ Math.\ Phys.\ \textbf{262}, 209 (2006).

\bibitem{Su1} A.~S\"ut\H o, Phys.\ Rev.\ Lett.\ \textbf{95}, 265501
    (2005).

\bibitem{Su2} A.~S\"ut\H o, Phys.\ Rev.\ B \textbf{74}, 104117
    (2006).

\bibitem{TS1} S.~Torquato and F.~H.~Stillinger,
    Exp.\ Math.\ \textbf{15}, 307 (2006).

\bibitem{PZ} G.~Parisi and F.~Zamponi, J.\ Stat.\ Mech.: Theory Exp.
    (2006), P03017.

\bibitem{TS2} S.~Torquato and F.~H.~Stillinger, Phys.\ Rev.\ Lett.\
    {\bf 100}, 020602 (2008).

\bibitem{CK} H.~Cohn and A.~Kumar, J.\ Am.\ Math.\
    Soc.\ {\bf 20}, 99 (2007).

\bibitem{ZST} C.~Zachary, F.~H.~Stillinger, and S.~Torquato,
    J.\ Chem.\ Phys.\ \textbf{128}, 224505 (2008).

\bibitem{CS} J.~H.~Conway and N.~J.~A.~Sloane, Discrete
    Comput.\ Geom.\ {\bf 13}, 383 (1995).

\bibitem{K} G.~Kuperberg, Geom.\
    Topol.\ \textbf{4}, 277
    (2000).

\bibitem{M} H.~L.~Montgomery,
    Glasg.\ Math.\ J.\ {\bf 30}, 75 (1988).

\bibitem{Mac} D.~MacKay, \textit{Information Theory, Inference, and
    Learning Algorithms\/} (Cambridge University Press, Cambridge, 2003).

\bibitem{SM} J.~F.~Sadoc and R.~Mosseri, \textit{Geometrical
    Frustration\/} (Cambridge University Press, Cambridge, 1999).

\bibitem{MK1} C.~D.~Modes and R.~D.~Kamien, Phys.\ Rev.\ Lett.\
    \textbf{99}, 235701 (2007).

\bibitem{MK2} C.~D.~Modes and R.~D.~Kamien, Phys.\ Rev.\ E \textbf{77},
    041125 (2008).

\bibitem{ZK1} P.~Ziherl and R.~D.~Kamien, Phys.\ Rev.\ Lett.\
    \textbf{85}, 3528 (2000).

\bibitem{ZK2} P.~Ziherl and R.~D.~Kamien, J.\ Phys.\ Chem.\ B
    \textbf{105}, 10147 (2001).

\bibitem{SDST} M.~Skoge, A.~Donev, F.~H.~Stillinger, and S.~Torquato,
    Phys.\ Rev.\ E \textbf{74}, 041127 (2006).

\bibitem{CCEK} H.~Cohn, J.~Conway, N.~Elkies, and A.~Kumar, Exp.\
    Math.\ \textbf{16}, 313 (2007).
\end{thebibliography}
\end{document}